\documentclass{sig-alternate}

\newcommand{\tran}{\textsl{R}}

\begin{document}
%
% --- Author Metadata here ---
\conferenceinfo{ACM RECSYS 2013}{'13 Hong Kong}

\title{Musical recommendations and personalization in a social network.}
\subtitle{[A case study]}

\numberofauthors{2} 

\author{
\alignauthor
Dmitry Bugaychenko\\
       \affaddr{Saint-Petersburg State University}\\
       \affaddr{Odnoklassniki LTD}\\
       \email{DmitryBugaychenko@gmail.com}
% 2nd. author
\alignauthor
Alexandr Dzuba\\
       \affaddr{Saint-Petersburg State University}\\
       \affaddr{Odnoklassniki LTD}\\
       \email{AlexandrDzuba@gmail.com}
}
\date{10 October 2013}

\maketitle
\begin{abstract}
This paper\footnote{This is a full version of a 4 pages article published at ACM RecSys 2013} presents a set of algorithms used for music recommendations and personalization in a general purpose social network \texttt{www.ok.ru}, the second largest social network in the CIS visited by more then 40 millions users per day. In addition to classical recommendation features like ``recommend a sequence'' and ``find similar items'' the paper describes novel algorithms for construction of context aware recommendations, personalization of the service, handling of the cold-start problem, and more. All algorithms described in the paper are working on-line and are able to detect and address changes in the user's behavior and needs in the real time.

The core component of the algorithms is a \emph{taste graph} containing information about different entities (users, tracks, artists, etc.) and relations between them (for example, user $A$ likes song $B$ with certainty $X$, track $B$ created by artist $C$, artist $C$ is similar to artist $D$ with certainty $Y$ and so on). Using the graph it is possible to select tracks a user would most probably like, to arrange them in a way that they match each other well, to estimate which items from a fixed list are most relevant for the user, and more.

In addition, the paper describes the approach used to estimate algorithms efficiency and analyze the impact of different recommendation related features on the users' behavior and overall activity at the service.
\end{abstract}

\category{}{Information Systems}{Personalization, Collaborative filtering, Social recommendation}

\terms{Algorithms, Case Study}


\section{Introduction}

The success of nearly all Internet projects nowadays highly depends on how much value can they provide for the user requiring as less efforts as possible. Recommender systems are one of the approaches used to increase ``value for effort'' rate. By analyzing user activity in the past a good recommender selects items which are most relevant to the current user's needs increasing user satisfaction, loyalty and the revenue of the project. Due to  the high demand for the recommender systems in the industry, there is a very large field of research works in this area.

The most commonly used recommenders are \emph{collaborative}~\cite{desrosiers2011comprehensive,koren2011advances} --- they analyze users behavior in the past and mine correlations between items and users. However, there is also a growing interest in the \emph{content based}~\cite{lops2011content} and \emph{social}~\cite{milicevic2010social} recommendation algorithms, in order to address limitation of the collaborative algorithms caused by the data sparsity and the ``cold-start problem''. It is now clear that none of these approaches is perfect and the best would be to combine benefits from all three worlds.

This paper proposes a composite approach for the recommender systems inspired by the \emph{random walk with restart} recommender from~\cite{konstas09rwr}. All data mined from the history of users' activity, content metadata and social network are combined in a \emph{taste graph}. Vertices of the graph represent different entities: users, tracks, artists, music genres, users' interests, etc. Edges of the graph represent relations between entities: a user likes a track, an artist is similar with another artist, an interest correlates with a track and so on. All edges are weighted according to the certainty of the relation and all weights are normalized in order to produce a \emph{stochastic system}.

The taste graph can be used to solve different kind of recommendation related task. In~\cite{konstas09rwr} authors propose to apply random walk with restart model in order to predict ratings for a user. Starting from the vertex representing the user process traverse the graph randomly preferring edges with higher weight and adding an additional \emph{restart probability} $\alpha$ (usually rather high) of restarting process from the initial vertex. \emph{Steady state probability} distribution of this process shows the strength of relation between the user and an item, which in many cases correlates with the probability that the user would like this item. However, this graph representation can be used in many other ways: given a set of items estimate their relevance for a user, identify groups of highly coupled items, extend a set of items with similar items (considering also user's preferences to them).

Taste graph based algorithms described in the paper are applied for the music service of ``OK'' (\texttt{www.ok.ru}) --- a general purpose social network started in 2006 with a goal to help people to find former classmates. Nowadays OK is the second largest social network in the CIS used mainly for fun and communication with rich media services (video, music, video chats and mail, etc.). Daily audience of the network is more than 40 millions users mainly from the Russian Federation, Eastern Europe and Middle Asia.

The paper is structured as follows. We start with the formal definition and the structure of the taste graph in section~\ref{sec:graph}. Section~\ref{sec:algorithms} describes basic algorithms on the graph and section~\ref{sec:improvements} addresses additional improvements made to provide better user experience. The impact of the recommender features on the service is analyzed in section~\ref{sec:impact} and final conclusions are made in section~\ref{sec:conclusion}.

\section{Taste graph}
\label{sec:graph}

In order to combine all the information mined from the system, including collaborative correlations, content information and social data a stochastic graph structure is used. Then this graph is analyzed by different algorithms in order to construct recommendations and personalize output. Formally, the taste graph is an oriented weighted labeled graph capturing all the information mined from the system. Formally taste graph $G$ can be defined as a tuple $\langle V, \theta, T_V, \tau_V, E, T_E, \tau_E, \tran, \omega \rangle$ where

\begin{itemize}
	\item $V$ is a finite non-empty set of \emph{vertices}, $T_V$ is a finite non-empty set of vertex types and $\tau_v: V \to T_V$ is a mapping function matching each vertex to its type.
	\item $\theta \in V$ is a \emph{zero balancing vertex} used to compensate impact of vertexes with small amount of outgoing edges.
	\item $E$ is a finite non-empty set of \emph{edges}, $T_E$ is a finite non-empty set of edge types and $\tau_e: E \to T_E$ is a mapping function matching each edge to its type.
	\item $\tran: E \to V \times V$ is a function matching each edge to its start vertex and end vertex. Edges leading to $\theta$ are \emph{zero balancing edges}.
	\item $\omega: E \to [0,1]$ is a weight function matching each edge to its weight.
\end{itemize}

A set of outgoing edges of a type $t \in T_E$ from a vertex $v \in V$ is defined as $out(v,t) \equiv \{e \in E \mid \tau_e(e) = t \wedge \exists v' \in V: \tran(e) = (v, v') \}$. Taste graph $G$ must satisfy following condition:

\begin{equation}
\forall v \in V, t \in T_E: \sum_{e \in out(v,t)}{\omega(e)} = 1.
\label{eq:limited_stochastic}
\end{equation}

Graph satisfying equation~\ref{eq:limited_stochastic} is called \emph{partly stochastic}. In order to get a fully stochastic graph balancing function $\beta: T_V \times T_E \to [0,1]$ is used. Balancing function must satisfy following condition:

\begin{equation}
\forall t_v \in T_V: \sum_{t_e \in T_E}{\beta(t_v,t_e)} = 1.
\label{eq:balancing_function}
\end{equation}

Given balancing function $\beta$ it is possible to define a \emph{balanced weight function} $\omega_{\beta} : E \to [0,1]$ as $\omega_{\beta}(e) \equiv \omega(e) \ast \beta(\tau_v(first(\tran(e))), \tau_e(e))$. It is clear that under weight function $\omega_{\beta}$ graph $G$ is a stochastic graph:

\begin{equation}
\forall v \in V: \sum_{t_e \in T_E, e \in out(v, t_e)}{\omega_{\beta}(e)} = 1.
\label{eq:full_stochastic}
\end{equation}

First of all the role of introduced non-standard constructs should be explained. Split of vertices and edges to different types and partial stochastic allows different parts of the taste graph to be constructed independently and then combined. For example, artists' similarity (edges of type ``artist $A$ is similar to artist $B$ with certainty $X$'') is constructed by a collaborative matching algorithm, while artists' tracks list (edges of type ``artist $A$ created track $B$ with weight $X$'') is constructed by an aggregate function from the overall track ratings. Furthermore, different parts of the graph can be updated at a different frequency, depending on the complexity of the update and the natural dynamics of the part.

Balancing function $\beta$ is used to manage impact of different factors on the overall result. For example, by increasing the weight of user--artist links we increase the impact of content-based recommendations, while increasing the user--track or user--user links would increase collaborative or social recommendations impact.

Zero balancing vertex $\theta$ plays an interesting role. In the taste graph the number of outgoing edges might vary from vertex to vertex. For example track $A$ might have $100$ similar tracks identified, while track $B$ only $50$. Smaller number of outgoing edges typically indicates the lack of statistics and low certainty of the list, but absolute weights on the edges from $B$ would be greater than on the edges from $A$ increasing its impact on the recommendations. In order to compensate this impact the zero balancing vertex is used. This vertex does not match any of the objects and has only a self-loop outgoing edge, but each time when the amount of outgoing edges of type $t$ for vertex $v$ is below limit, an edge from $v$ to $\theta$ added to ``drain'' weights from the existing edges. The weight of the zero balancing edge is defined in a way that it approximates the sum of weights of missing edges. For collaborative similarity this weight is typically approximated by a \emph{linear decay}, while for overall popularity \emph{exponential decay} is used.

As of now the taste graph for musical recommendations consists of the following parts:

\begin{description}
	\item[Users' preferences,] encoding the relations between users and the items they like (tracks and artists).
	\item[Artists' similarity,] based on the collaborative correlations between artists in the music catalog.
	\item[Tracks' similarity,] capturing the collaborative and temporal correlations between tracks in the music catalog.
	\item[Artists' tracks,] representing a set of tracks produced by an artist weighted by overall popularity.
\end{description}

Users' preferences is the most dynamic component of the graph --- it is updated each time preferences are requested. Preferences, in turn, are composed from different parts. All the history of user playbacks is recorded into a data warehouse and aggregated at a daily basis. Furthermore, there is a running window of the 1000 last played tracks with the time stamps in a real-time storage which is also considered. In addition to the playbacks history there is a real-time storage of the users' custom play lists to consider.

The artists' similarity is created by an analysis of playbacks history. Analysis is done in multiple steps: first of all from a set of all artist's tracks representative tracks are selected, then playbacks for these tracks are aggregated in order to get user--artist matrix. Using a simple correlation measure artist--artist matrix is constructed and then more precise correlation measure is applied iteratively to artist--artist matrix to refine it. Finally, outliers filtration algorithms are applied to the matrix and for each artist the top similars are selected.

For the tracks' similarity matrix for each pair of tracks $A$ and $B$ an amount of playbacks from the same user in the scope of limited time window is counted. Than the baselines are applied in order to address tracks popularity and results are normalized. In the end outliers filtration is applied and the top similars are selected.

The artists' works are selected from the music catalog meta data which keeps mapping from track to its main artist and overall rating of the tracks in the system is used as a weighting function. Track ratings are analyzed in dynamics in order to put tracks with higher interest recently above others, even if they have lower overall rating. Additional boost is applied for fast growing tracks introduced recently.

\section{Algorithms}
\label{sec:algorithms}

Based on the data of the taste graph algorithms for solving most of the standard recommender tasks can be constructed.

\subsection{Generating recommendations}
\label{subsec:rwr}

This task can be formulated as follows: ``given a user $u$ select a set of items $I$ such that the user would most probably like''. In~\cite{konstas09rwr} a random walk with restart model is proposed for this task. The main idea of the algorithm is to model a process of random graph traversing starting from the user's node with the probability $\alpha$ to restart at each step. The state of the process is represented by a vector $x \in [0,1]^V$ capturing for each vertex $v \in V$ the probability of being in this vertex. Given the current state $x$ next state $next(x)$ is calculated as follows:

\begin{equation}
next(x) \equiv \alpha \ast |u| + (1 - \alpha) \ast \sum_{v \in V}{\left(x_v \ast next(v)\right)},
\label{eq:rwr}
\end{equation} 
where $u \in V$ is the initial vertex corresponding to the user, $|u| \in [0,1]^V$ is a vector with a single $1$ for the vertex $u$ and $0$ for all other nodes, and $next(v)$ is a vector calculated as follows:
\begin{equation}
next(v) \equiv \sum_{t_e \in T_E, e \in out(v, t_e)}{\omega_{\beta}(e) \ast |second(\tran(e))|}.
\label{eq:next_v}
\end{equation}

Given initial vertex $u$ random walk recommender finds a fixed point $x_f$ of the operation $x = next(x)$ also known as the \emph{steady-state probability distribution}. Items with higher steady-state probability are most likely to be relevant for the user, thus selecting top $N$ items from the vector should be a good choice of recommendations.

One of the problems with RWR approach is that the items from the initial user's preferences get higher probability comparing to non-rated items and it makes recommendation useless if not addressed (the user would see only items he already knows well). In order to address this the computation is split in two stages: calculating next state without applying restart $x = \sum_{v \in V}{\left(x_v \ast next(v)\right)}$ and then restarting $x = \alpha \ast next(u) + (1 - \alpha) \ast x$. In the end of the process state vector \emph{before} restart is used as a result. Additionally, extra suppression can be applied to the already known items (up to their full exclusion).

\subsection{Personalizing list}
\label{subsec:personalization}

Another common goal for a recommender is to select most relevant items for user $u$ from a set of items $I$ (which could be, for example, a response of a search engine or the set of top tracks). With the taste graph this could be done by traversing the graph from user preferences vector in a limited number of steps counting all the visits of the vertices from target vector. The vertices with higher number of more probable visits are considered as more relevant for the user.

Formally, having the initial state vector $x_0 = next(u)$, the target vector $x_t$ and the weight vector $w = (w_0, \ldots, w_n)$ the relevance of the items in $x_t$ according to $x_0$ is calculated as follows:

\begin{equation}
rel(x_t) \equiv x_t \ast w_n + \sum_{i \in \{0,\ldots, n - 1\}}{w_i \ast x_i|_{x_t}},
\label{eq:personalization}
\end{equation}
where $x_i = next(x_{i-1})$ for $i > 0$ and $x_i|_{x_t}$ denotes projection of the vector $x_i$ on the non-zero components of the vector $x_t$. Weights $w_0$ and $w_n$ play a special role in this scheme: $w_0$ is used to weight the intersection of target and source is estimated and $w_n$ represents the impact of original weights in the target vector.

\subsection{Extending list}

In order to help users to find music they like with less efforts it is worth  introducing a feature ``extend list''. The list could be a user's custom play list, any play list the user is browsing now, or even a single-item list with a song the user is listening now or an artist the user is looking at.

The same random walk technique described at~\ref{subsec:rwr} can be used to identify items similar to the given initial list and personalization technique described at~\ref{subsec:personalization} can be used to tune results using user's preferences. For a too small or a too unvaried list the input can be enriched by appending other items from the user's preferences which are highly coupled with the initial list. The same personalization technique or a more advanced tuning technique (described at~\ref{subsec:tuning}) can be used to identify those items.

\section{Additional improvements}
\label{sec:improvements}

Section~\ref{sec:algorithms} provides a good foundation for a recommender system, however several improvements are required in order to achieve good results in practice. The most important ones are described in this section.

\subsection{Randomizing recommended sequence}

Most works on the recommender algorithms are ended by ranking items for a user, however in the music radio case it is not enough. Each time user opens the radio he would like to get a new sequence which is

\begin{description}
	\item[Relevant,] all items have high predicted rating.
	\item[Coherent,] items are well listened together (have high similarity level).
	\item[Diverse,] songs from many different artists are included.
	\item[Novel,] includes tracks and artists user has not listened to and is unlikely to find them by himself.
\end{description}

Returning a head with the top rated items provides user with relevant sequence (assuming recommender has a reasonable performance). Introducing Gaussian noise or picking random items from the weighted list looks like a solution, but actually none of these provides desirable quality. A better approach could be to use \emph{random pick with rejection}. Items are picked randomly from the weighted list and each selected item is challenged by a probabilistic rejection algorithm, which depends on multiple factors.

Given a vector of predicted preferences $x \in [0,1]^V$ constructed by the algorithm described at~\ref{subsec:rwr} and a \emph{strict ordering} of the graph nodes\footnote{Exact selection of the ordering is not important, but it is required to be strict and total.} $\succ \subset V \times V$ it is possible to construct vector $x_{\succ} \in [0,1]^V$ such that $x_{\succ}[v] = x[v] + \sum_{v' \in V, v' \succ v}{x[v']}$. Using vector $x_{\succ}$ random pick can be implemented using a uniformly distributed over $[0,1]$ random variable $r$ and picking the closest upper bound from the $x_{\succ}$ with binary search. 

When item $v$ is picked, it is challenged by a family of \emph{rejection factors} $\{rf_i: V \times V^* \to [0,1]\}_{1}^{N_{rf}}$, where each factor $rf_i$ given an item to check and a sequence of already selected items defines the probability of the item to be kept. The overall probability for the item $v$ of being added to the sequence $S \in V^*$ is $p_a = \prod_{1}^{N_{rf}}{rf_i(v, S)}$ and the item is added to the sequence if the next value of $r$ is less than $p_a$.

Among the rejection factors the most important is the \emph{previous presence} of the item and other items from the same artist. Presence of item itself can be covered by a simple ``yes or no'' factor prohibiting repeated inclusion of the item either for the entire sequence or in the tail of last $N$ items in the sequence. Presence of other works from the item's artist can be considered in two dimensions: the overall presence and the distance from the previous entry. The overall presence is modeled as an exponentially descending factor $rf(v, S) = \beta^{counta(v, S)}$ where $\beta \in [0,1]$ is the descend speed and $counta: V \times V^* \to \{0,1, \ldots\}$ is a function returning amount of items created by the artist of $v$ in the sequence $S$. The previous entry factor is modeled as inversed exponential descend $rf(v, S) = 1 - \gamma^{tailposa(v,S)}$ where $\gamma$ is the speed of descend and $tailposa: V \times V^* \to \{0,1, \ldots\}$ is the first position of the work from the artist of $v$ in the tail of $S$.

While previous presence rejection factors increase diversity and novelty of the recommended sequence, coupling with nearest predecessors factors increase coherence of the sequence. The main idea here is to estimate the weight of the paths between item $v$ and its nearest predecessors in the sequence using the taste graph which could be done by using the personalization technique described at~\ref{subsec:personalization}.

\subsection{Context awareness}
\label{subsec:tuning}

After the first experiments with music recommendations it was recognized that a single user might have preferences of very different kinds and these preferences, if considered as a whole, produce relevant, but useless recommendations. In order to provide value for the user, recommended music must address current context the user is in now (working, having a dinner or playing with his kids). The problem of context awareness is explored in~\cite{adomavicius2011context} and two techniques were adopted to generate contextual recommendations:

\begin{description}
	\item[Contextual pre-filtering] of the user preferences --- only items matching current context are selected for expansion.
	\item[Contextual post-filtering] of the generated recommendations --- items having to weak linkage with the context are removed from the result. 
\end{description}

The filtering is done by analyzing all paths of length $N$ starting from the \emph{context set} and ending in the vertices of \emph{filtered set}. The analysis is done in the same way as described in~\ref{subsec:personalization}. Three measures are collected: count of the paths, sum of weights on the all paths and the weight of the best path. Vertices reached high enough limit for at least one of the measures are selected as the input for recommendations generation (when doing pre-filtering) or randomization algorithm (when doing post-filtering).

The key factor for efficient contextualization is a selection of the proper context set. The selection is done by the user himself, but he is provided with a bundle of automatically generated context sets to select from. Each set contains items from the user's preferences which are worth listening together, thus being used as a context set they produce coherent recommendations.  \emph{User preference clustering} is done behind the scene to create user specific context sets.

Using the taste graph a set of clustering algorithms based on the identification of \emph{connected components} was implemented. The simplest approach is \emph{weight bound} clustering. In this case given the weight limit $\tau \in (0,1)$ and a vector of user's preferences $x_u \in [0,1]^V$ \emph{user preferences graph} $G_u = \langle V_u \subset V, E_u \subseteq V_u \times V_u \rangle$ is constructed such that
\begin{eqnarray}
V_u & = & \{v \in V \mid x_u[v] > 0\}, \\
E_u & = & \{(v,v') \in V_u \times V_u \mid v' \in nbr_{\tau}(v) \},\
\label{eq:weightboundclustering}
\end{eqnarray}
where $nbr_{\tau}(v) \equiv \{v'' \in V \mid next(v)[v''] \geq \tau \}$ is the set of direct neighbors of $v$. Linked components $\{C_i\}_1^*$ of the graph $G_u$ are used as the preferences clusters.

Better results can be achieved by using \emph{commons bound} clustering. In this case two limits are defined: weight limit $\tau \in (0,1)$ and common count limit $nc \in \{1, 2, \ldots \}$, and the edges of graph $G_u$ are constructed as follows:
\begin{equation}
E_u = \{(v,v') \in V_u \times V_u \mid \left\| nbr_{\tau}(v) \cap nbr_{\tau}(v'))\right\| \geq nc \},\
\label{eq:commonboundclustering}
\end{equation}
where $\left\| nbr_{\tau}(v) \cap nbr_{\tau}(v'))\right\|$ is the amount of common neighbors for $v$ and $v'$. This approach decreases influence of bridges in the taste graph producing more coherent components. 

However, none of the approaches can handle the diversity of the users' preferences with single settings --- as long as user listens to more and more items, more and more bridges rise in his preferences causing them all to merge in a single component. In order to handle preferences of different size well hierarchical clustering were applied. Clustering algorithm with relaxed limits produces initial separation and clusters above the size limits are passed to a more tightly configured algorithm.

Initially the system was lunched with a chain of 7 clustering algorithms based on common bounds clustering and the users feedback revealed two important problems:
\begin{description}
	\item[Low coverage] due to too many items being separated into a single-item clusters.
	\item[High dynamics] caused by a fast growth of the clusters causing them to be passed to a more tight clustering algorithm. 
\end{description}

Which is more important, user's favorite items tend to cut off into a small cluster more often since they usually have a more diverse neighborhood sets. And the same is true for user's favorite context sets which grow fast. In order to reduce the speed of the process a relative limit was imposed for the selection of items (only items with the rating above $x\%$ of the average rating are considered when clustering) and \emph{fallback clustering} (passing the set of rejected items to the same clustering algorithm) was used. These techniques increased the coverage and stability, but results still were unsatisfactory.

In order to solve the problem a more high-order similarity measure based on \emph{common neighborhood subgraph density}~\cite{kang2009common} and a new clustering algorithm based on \emph{affinity propagation}~\cite{dueck2009affinity} were applied. Having a set of user's preferences $V_u \subset V$ we define the similarity function $s: V_u \times V_u \to \mathbb{R}$ such that
\begin{eqnarray}
s(v,v')  \equiv 
\left\| \left\{e \in E \mid R(e) \in \left(nbr_0(v) \cap nbr_0(v')\right)^2 \right\} \right\|.
\label{eq:common_neighborhood_density}
\end{eqnarray} Intuitively this similarity measure counts edges in the taste graph with both ends belonging to the $v$ and $v'$ common neighborhood. The self-similarity of an item $s(v,v)$ in this case is not less than similarity with any other item, which causes affinity propagation to create a lot of small clusters. To avoid that, the self-similarity is multiplied by a discount factor $\delta \in [0,1]$.

Affinity propagation clustering~\cite{dueck2009affinity} is a relatively new clustering technique based on the simulation of message passing between vertices. For each pair of nodes $i,j \in V_u$ \emph{responsibility} $r(i,j) \in \mathbb{R}$ and \emph{availability} $r(i,j) \in \mathbb{R}$ are defined. Responsibility, $r(i,j)$, is a message from $i$ to $j$ that reflects the accumulated evidence for how well-suited $j$ is to serve as the \emph{exemplar} for $i$. Availability $a(i,j)$ is a message from $j$ to $i$ that reflects the accumulated evidence for how appropriate it would be for $i$ to choose $j$ as its exemplar. Initially all availabilities and responsibilities are set to $0$ and then updated using the following schema:

\begin{eqnarray*}
r  & = &  (1 - \lambda) \cdot p + \lambda \cdot r \\
p(i,j)  & = & \left\{
  \begin{array}{ll}
	  s(i,j) - \max\limits_{k \neq j}(a(i,k) + s(i,k)) & i \neq j \\
	  s(i,j) - \max\limits_{k \neq j}(s(i,k)) & i = j 
  \end{array}\right. \\
a  & = &  (1 - \lambda) \cdot \alpha + \lambda \cdot a \\
\alpha(i,j)  & = & \left\{
  \begin{array}{ll}
	  \min(0,r(j,j) + \sum\limits_{k \neq i,j}\max(0,r(k,j)) & i \neq j \\
	  \sum\limits_{k \neq i}\max(0,r(k,j))& i = j 
  \end{array}\right.
\label{eq:affinity_propagation}
\end{eqnarray*} Where $p$ and $\alpha$ are propagating responsibility and availability respectively and $\lambda$ is the damping factor. Having responsibilities and availabilities calculated exemplars for vertices are selected as:
\begin{equation}
ex(i) \equiv \arg \max\limits_{j \in V_u} (r(i,j) + a(i,j)).
\label{eq:exemplars}
\end{equation} Algorithm is repeated until convergence of $a$ and $r$ or, which is more often in practice, until stabilization of $ex$ for several iterations (\emph{convince limit}). Vertices selected as a candidates for themselves ($ex(i) = i$) are considered as the centers of the clusters and a cluster $C_i$ for center $i$ is extracted as follows:
\begin{equation*}
C_i \equiv \left\{ j_0 \in V_u \mid \exists j_1, \ldots, j_n \in V_u: ex(j_0) = j_1 \wedge \ldots \wedge j_n = i \right\}
\label{eq:}
\end{equation*}

Comparing to the linked component extraction affinity propagation is more computationally intensive, but it produces more stable and relevant results. Furthermore, similarity measure based on the common neighborhood density is more computationally intensive by itself then the common neighbors count and the direct link weight. In order to reduce computations and still support on-line context set generation similarity values are calculated once and cached in memory. Further improvement can be achieved using pruning technique described at~\cite{fujiwara2011fastaffinity} to remove edges which do not influence the result from the search.

Affinity propagation technique handles well users with large history and genres with high coherence (classical music, child songs, rock), however for users with small preferences belonging mainly to vague genre (Russian rap is the worse) it tends to produce too small clusters. In order to compensate this effect user preferences are enriched (see~\ref{subsec:coldstart} for details) and small clusters are relinked to keep the high coverage level.

\subsection{Addressing cold-start problem}
\label{subsec:coldstart}

Cold-start problem can be seen from two perspectives:

\begin{description}
	\item[New user] joins the system without enough information to provide him with reliable recommendations.
	\item[New item] added without known reliable links with other items, while it might be the most interesting item for the users due to its novelty.
\end{description}

In both cases there are not enough collaborative information related to the user and the item to make conclusion, however, there are content-based and system-wide information which can be used.

When a new user joins the system it is important to provide him with some recommendations --- showing a message like ``find something yourself and get back later'' is not a good idea (in most cases user won't get back). To support new users demographical information is used. All users are split into segments depending on their age, sex and region. Behavior of the users in the scope of a demography group is analyzed in order to create a \emph{demography profile}. When a user has no own preferences, best matching demography profile is used instead. Even if user already has some preferences, but not that much, then his own preferences are mixed with demography profile.

When new item is added to the system it becomes the most interesting candidate for being recommended. However, due to the lack of statistics it has only a few links with other items and low overall rating, thus it is unlikely to be recommended. Handling of new items includes two stages:

\begin{description}
	\item[Identify] interesting new items (not every new item is really interesting).
	\item[Boost] identified items in the recommendations and the top list for relevant users.
\end{description}

Even for a brand new item there is some activity in the system which can be used to estimate its relevance, but this activity could be way below activity around well known old hits. In order to make relevant new items more visible, only items recently added to the system are considered and activity around them are compared to the time they spent in the system. For item $v \in V$ relevance $rel(v)$ can be estimated as follows:
\begin{equation}
rel(v) \equiv \frac{interested\_users(v)}{days(today - date(v) + 1)^{ti}},
\label{eq:relevance}
\end{equation} where $interested\_users(v)$ is the amount of distinct users listened the item (or added the item to a playlist), $date(v)$ is the date when item had been added and $ti \in (0,+\infty)$ is the \emph{time impact factor}. Items with $rel(v)$ above novelty limit are considered novel and boosted in the recommendations.

When using random walk technique for recommendation generation, items with the highest sum of weight on the incoming edges has (in general) higher probability of being recommended. Old items usually have many links between each other, but not with new items. The only reliable link in the graph leading to the new item is the artist--track link and in order to boost item in the recommendations the weight of this link is set to a very high level comparing to other links from this artist. Furthermore, the overall system wide rating of the item is also boosted in order to consider it while creating a personalized content of the main page for the users (see~\ref{subsec:personalization}).

This technique works well for boosting new tracks made by a well-known artists (according to the author personal experience new tracks becomes visible in 24 hours after their introduction to the system), but it does not handle case when a new potentially interesting artist is added. In order to support this case more high level properties (for example genre) needs to be analyzed. However, introduction of more high level concepts in the graph can decrease relevance of the recommendations (musical genres can be very vague).

\section{Analyzing impact}
\label{sec:impact}

Evaluating a recommender system is not a trivial process by itself (see~\cite{shani2011evaluating} for details). One of the most common approaches to evaluation is the offline analysis based on the historical data. Information about user behavior is split into training and test set, recommender is configured using training set only and then its predictions compared with the observed ratings in the test set. This approach is simple and cheap, however it has an inherent drawback --- recommender proposing items user probably knows always get higher score then the recommender proposing relevant items user not likely knows and can't find them himself.

An alternative to offline analysis is an online experiment either on a selected subgroup of the users, or on all of them. Online experiments can be very informative, but they require significant time, can lead to a negative impact on the real users and their results can be influences by many other factors not controlled (and some times even not known) by the experimenter. Despite of the all difficulties related to the online experiments, they are considered as the best approach for impact analysis and there is a rich and flexible back-end for statistics collection and analysis implemented in order to support them.

\begin{figure*}
\centering
\psfig{file=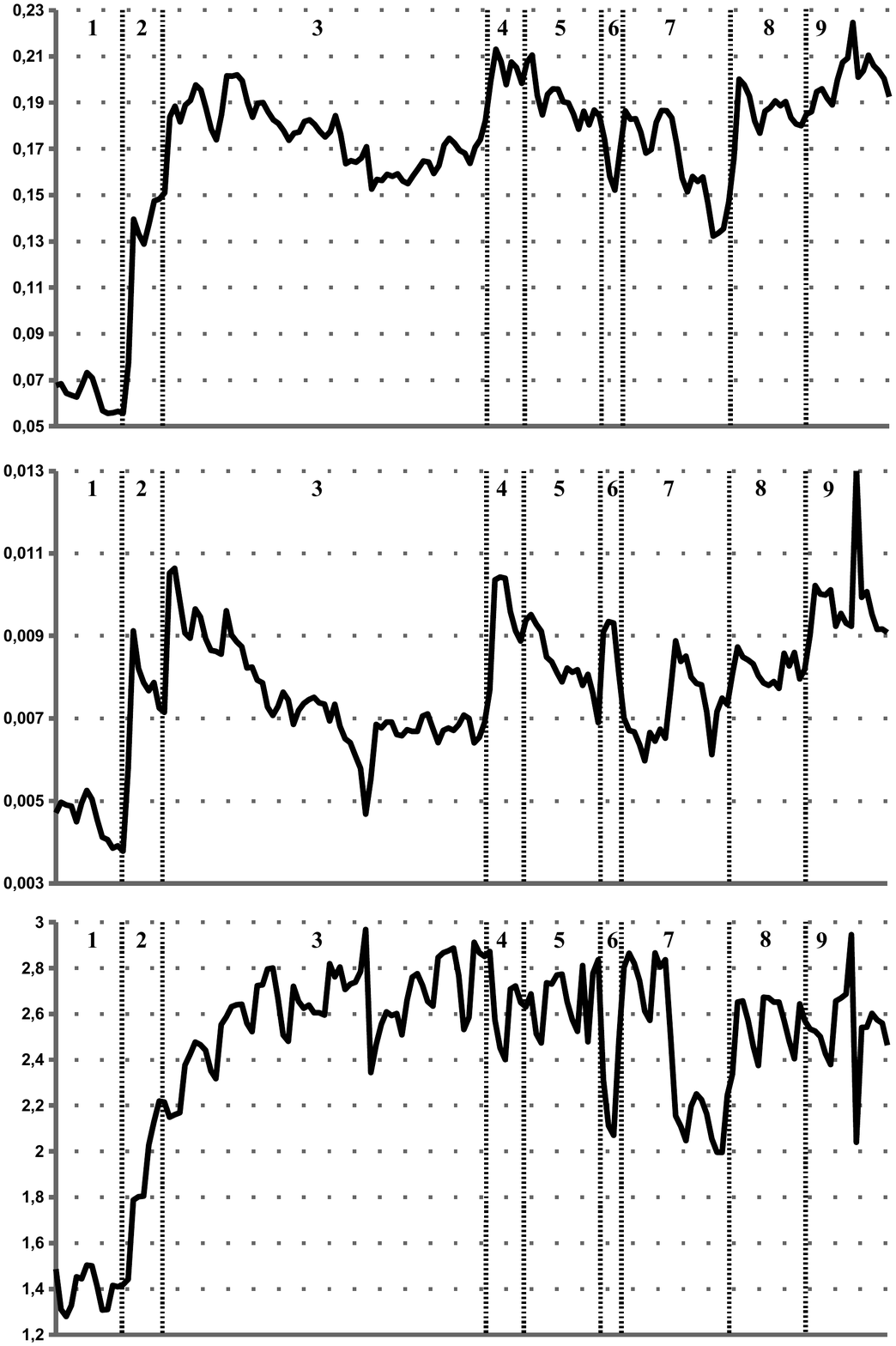,height=0.95\textheight}
\caption{Playback and ``likes'' from the main page comparing to playbacks from ``My music'', playbacks from the main page comparing to clicks.}
\label{fig:graphs}
\end{figure*}

Personalization of the main page (the list of 100 songs user sees after entering the music area) was the first significant feature based on the recommender launched at 2012-11-16. Figure~\ref{fig:graphs} shows user activity trends during the 6 months passed after the launch. Three main indicators are collected:
\begin{enumerate}
	\item Amount of playbacks from the main page comparing to playbacks from ``My music''.
	\item Amount of ``likes'' from the main page comparing to playbacks from ``My music''.
	\item Amount of playback from the main page comparing to the amount of clicks at the main page.
\end{enumerate}
``My music'' is the list of the user's favorite tracks, it is the most popular list in the system and is used as the measure of overall level of activity. In this perspective the first indicator shows how relevant is the content at the main page for the users, the second indicator shows how novel is it and the third shows how confident the users are (how often they have to control playbacks manually). There are nine periods identified corresponding to the changes in the personalization algorithm:

\begin{enumerate}
	\item Main page contains top tracks based on the all users activity in the system for last 30 days.
	\item Personalization enabled selecting top 100 tracks for a user among 1000 top tracks in the system.
	\item Size of the underlying list increased to 1500.
	\item Enabled suppression of the known items ($w_0$ set to negative value).
	\item Changed the underlying list preparation algorithm.
	\item Track known by a user removed from the main page.
	\item Full removal disables, Gaussian noise introduced into personalization.
	\item Strength of the known tracks suppression decreased, underlying list size increased to 3000.
	\item Use of demography data enabled for the new users.
\end{enumerate}

According to the first two charts at the figure~\ref{fig:graphs} personalized main page performs two-three times better then the non-personalized variant. During initial experiments (stages 2-5) each time settings of the personalization were adjusted the user activity at the main page increased (users seen new content there), but a downward trend emerged after few days (users got used to what they see). Stage 6 showed that the strong filtering of known content increases novelty (amount of likes increased), but relevance and the users' interest decreased (playbacks dropped). Replacement of the strong filtering with a lighter suppression (known content still might appear, but with a decreased weight) and introduction of the Gaussian noise at the stage 7 changed trend for likes to upward and for playbacks to sideways. Increasing  the underlying list size and using demographical data we achieved upwards trends in both areas.

The third chart at the figure~\ref{fig:graphs} shows the evolution of the users' confidence in the content of the main page. For lists users are confident in they tend to turn playback on with a click and listen, but for the lists with low confidence they tend to control playbacks thoroughly (producing more clicks). With the personalization enabled the users' confidence increased slowly and without downward trend. Attempts to increase novelty (at stages 6 and 7) increased the amount of clicks, but it is more an indication of the increased curiosity then of the decreased confidence. The confidence chart also reveals weekly trend --- at weekends users tend to produce more clicks.
 
\section{Conclusions and further work}
\label{sec:conclusion}

Recommender systems and personalization are a crucial tool for nearly all kind of web sites. For \texttt{www.ok.ru} introduction of personalization for the content of the main page in the music area has increased user activity there threefold. Furthermore, it created a driver for the entire service. This is a good example of the fact that the common recommender's use case ``recommend a sequence'' is not the only interesting case. Thus, modern industrial recommender must be flexible enough to address different tasks (recommend a sequence, extend a list, personalize a list, etc.) based on the common foundation.

On the others hand, it is clear that in order to provide high quality recommendations there is a need to consider the information of the different kinds: collaborative correlations, content information, context, demography, etc. Orchestrating multiple recommenders can address this challenge, but at the cost of increased computation complexity (architectural and deployment complexity increases to). Thus, again, there is a need for a common foundation capable of incorporating data of different kind.

Taste graph described in this paper has many of the desired properties: it can include data of different kind in a single model with online balancing, different kind of algorithms for many tasks can be implemented, computational complexity can be made reasonable for online computations and immediate reaction to the changes in the user behavior. The results achieved at \texttt{www.ok.ru} are very promising in terms of both increased user activity and the recommender's non-functional properties (computational efficiency, scalability and flexibility).

We are going to continue research and experiments with the taste graph in multiple directions: 

\begin{description}
	\item[Demographical adjustments.] Some collaborative correlations included in the graph are enforced by the data only from certain demography group and might be irrelevant for others.
	\item[User personal adjustment.] Even if a correlation is system wide, it still might be irrelevant for a certain user.
	\item[Integration of other collaborative data.] So far collaborative data in the graph are neighborhood based, but introduction of SVD factors, SVM classes, etc. might increase performance.
	\item[Cluster analysis.] Clustering technique use for the context sets generation being applied for the larger sets is also a promising approach for filtering and enriching graph.
\end{description}

Furthermore, it is worth to consider application of the taste graph based recommender platform for other types of content at \texttt{www.ok.ru} (games, communities, gifts, videos, etc.).

\end{document}